\documentclass[a4paper,11pt]{article}

\usepackage{lineno}

\usepackage[utf8]{inputenc}

\usepackage{hyperref} 
\usepackage{url}
\usepackage{graphicx}
\usepackage{amssymb}
\usepackage{amsmath}
\usepackage{multirow}

\usepackage{graphics}
\usepackage{natbib}
\graphicspath{{./FigureProc}}
\usepackage{color}

\usepackage{xcolor}
\usepackage{textcomp}

\usepackage{orcidlink}

\usepackage{comment}
\usepackage[blocks]{authblk}

\title{Test beam performance of a novel RICH detector with timing capabilities for the future ALICE~3 PID system at LHC}

\author[a,1]{M.~N.~Mazziotta\,\orcidlink{0000-0001-9325-4672}}
\author[a]{L.~Congedo\,\orcidlink{0000-0003-4536-4644}}
\author[a]{G.~De~Robertis\,\orcidlink{0000-0001-8261-6236},}
\author[b]{A.~Di~Mauro\,\orcidlink{0000-0003-0348-092X},}
\author[a]{F.~Licciulli\,\orcidlink{0000-0002-6955-0321},}
\author[a,c]{L.~Lorusso\,\orcidlink{0000-0002-2549-4401},}
\author[b]{P.~Martinengo\,\orcidlink{0000-0003-0288-202X},}
\author[a]{E.~Nappi\,\orcidlink{0000-0003-2080-9010},}
\author[a,c]{N.~Nicassio\,\orcidlink{0000-0002-7839-2951},}
\author[a,c]{G.~ Panzarini\,\orcidlink{0000-0002-2586-1021},}
\author[a,c]{R.~Pillera\,\orcidlink{0000-0003-3808-963X}}
\author[a,c]{and G.~Volpe\,\orcidlink{0000-0002-2921-2475}}

\affil[a]{Istituto Nazionale di Fisica Nucleare (INFN), Sezione di Bari, \\ via Orabona 4, I-70126 Bari, Italy}
\affil[b]{CERN, the European Organization for Nuclear Research, \\ Esplanade des Particules 1, 1211 Geneva, Switzerland}
\affil[c]{Dipartimento di Fisica dell'Universit\`a e del Politecnico di Bari, \\ via Amendola 173, I-70126 Bari, Italy}

\affil[1]{Corresponding author: mazziotta@ba.infn.it}

\date{}

\begin{document}
\maketitle

\begin{abstract}
The ALICE Collaboration is proposing a completely new apparatus, ALICE 3, for the LHC Run 5 and beyond. A key subsystem for charged particle identification will be a Ring-Imaging Cherenkov (RICH) detector consisting of an aerogel radiator and a photosensitive surface based on Silicon Photomultiplier (SiPM) arrays in a proximity-focusing configuration. A thin high-refractive index slab of transparent material (window), acting as a second Cherenkov radiator, is glued on the entrance face of the SiPM arrays to achieve precise charged particle timing. Requiring time matching between aerogel Cherenkov photon and track hits leads to an improvement of pattern recognition by discarding the uncorrelated SiPM dark count hits.

In this work we present the current status of the R\&D performed for the ALICE 3 RICH detector prototype and the expected full scale system performance. A special focus will be given to the beam test results obtained with a small-scale prototype instrumented with various array of Hamamatsu SiPMs with pitches ranging from 1 to 3 mm. The Cherenkov radiator consisted of a 2 cm thick aerogel tile with a refractive index of 1.03 at 400 nm  wavelength. For timing measurements SiPM arrays coupled with two different window materials (SiO$_2$ and MgF$_2$) were used. The prototype was successfully tested in beam test campaigns at the CERN PS T10 beam line.
The data were collected with a complete chain of front-end and readout electronics based on the Petiroc 2A and Radioroc 2 together with a picoTDC to measure charges and times. We measured a charged particle detection efficiency above 99\% and a single photon angular resolution better than 4.2 mrad at the Cherenkov angle saturation with a time resolution better than 70 ps for charged particles.
\end{abstract}



\section{Introduction}
\label{sec:intro}

The ALICE Collaboration is proposing a completely new apparatus, ALICE 3 ~\cite{ALICE3Loi}, for LHC Runs 5 and 6.
A key requirement for achieving the ALICE 3 physic goals is an extensive particle identification (PID) of $e^{\pm}$, $\mu^{\pm}$, $\pi^{\pm}$, $K^{\pm}$, $p$ and $\bar{p}$ with different dedicated subsystems providing acceptance over eight units of pseudorapidity ($|\eta|<4$). The design target of the ALICE 3 PID system is to ensure a better than 3$\sigma$ $e$/$\pi$, $\pi$/$K$ and $K$/$p$ separation for momenta up to \mbox{2 GeV/$c$}, \mbox{10 GeV/$c$} and \mbox{16 GeV/$c$}, respectively.

The required performance in the barrel RICH (bRICH) is achieved using a radiator with a refractive index $n~=~1.03$ at 400 nm wavelength and requiring an overall ring angular resolution better than 1.5~mrad at Cherenkov angle saturation.
The required resolution is expected to be achieved using a proximity-focusing layout ~\cite{Eugenio_RICH}, with 2 cm thick aerogel radiator tiles separated by an expansion gap larger than 20 cm from a photosensitive surface.
Since the bRICH will operate in a 2 T solenoidal magnetic field, the most promising currently available photon sensor technology is represented by SiPMs, compared to, for example, vacuum-based devices such as  Microchannel Plate Photo-Multiplier Tubes (MCP-PMTs) and Large Area Picosecond Photodetectors (LAPPDs), which suffer from the presence of such magnetic field configuration.
The photodetector surface will be segmented in $2\times2$~mm$^2$ cells providing a single photon detection efficiency above 40\% at wavelengths close to 400 nm. 
As a further requirement, a single photon time resolution of $\mathcal{O}(100~\text{ps})$ or better is needed to disentangle signal photon hits from background ones. 
In this work we report the status of the R\&D activities focusing on the beam test results obtained with various RICH prototypes~\cite{rich1}.

\section{ALICE 3 RICH prototypes}
\label{sec:richprototype}

RICH prototypes have been successfully assembled and tested at the CERN-PS T10 beam line since 2022~\cite{Nicassio_proceeding_IWASI,Mazziotta_IWORID}.
All tested  prototypes consisted of cylindrical vessel housing a Cherenkov radiator and a photodetector layer. The radiator, made of a 2 cm thick hydrophobic aerogel tile from Aerogel Factory \& Co. Ltd. with a refractive index of $n$ = 1.03 at 400~nm wavelength~\cite{AnnaRita_proceeding_ICHEP}, was located 23~cm upstream with respect to the photodetector layer, which was equipped with SiPM arrays. 
The SiPM arrays were connected to custom printed circuit boards (PCBs) mounted on copper plates and cooled down to -5$^\circ$C using water chillers and Peltier cells. The vessel was flushed with Ar to keep the relative humidity~$<2\%$. Temperature sensors TT4-10KC3-T125-M5-500~\cite{ntc10k} were assembled on the copper plates to monitor their temperatures by using Raspberry Pi 3 with ADS1115 16-Bit ADCs \cite{ads1115}. In addition, humidity sensors SHT31-D \cite{sht31} were used to monitor the room and the vessel humidity.
 
The analog SiPM signals were read-out by custom front-end boards (FEBs) by means of 1 meter long high speed 50 Ohm multi-channel Samtec HLCD-20 (40 channels) cables~\cite{hlcd}. A feed-through PCB connector was used to route the analog signals from inside the vessel to the FEBs placed outside.
All the SiPM PCBs were equipped with 1-Wire digital temperature sensors DS18B20~\cite{ds18b20} to monitor their temperatures (the 1-Wire with the 3.3V voltage bias signal was still routed by means of a HLCD channel). Finally, the SiPM bias voltages were provided by means of 4 HLCD channels.
As for charged particle tracking and for beam monitoring, the setup included two X-Y tracker modules based on plastic scintillating fibers coupled with HPK S13552 linear SiPM arrays (128 channels with $0.23\times1.625$ mm$^2$ effective area) mounted outside the vessel  and are operated at room temperature \cite{MAZZIOTTA2022167040,Roberta_proceeding_IWASI}.

In the first version of the prototype (2022 and 2023), the Cherenkov-photon detector consisted of eight HPK S13552 SiPM arrays~\cite{MAZZIOTTA2022167040} with centers arranged along a circumference of 6.5 cm of radius. Two additional HPK S13361-3075AE-08 SiPM arrays (8$\times$8 channels with 3.2 mm of pitch), hereafter called M0 and M1, were also placed in the vessel along the beam line to perform charged particle timing measurements~\cite{Mazziotta_IWORID}. The M0 array was located 10 cm upstream with respect to the aerogel tile, while the M1 array was placed in the center of the Cherenkov-photon detector.
A 1 mm thick MgF$_{2}$ ($n= 1.38$ at 400~nm) window was glued on M0, while a 1 mm thick SiO$_{2}$ ($n = 1.47$ at 400~nm) radiator was glued on M1. The 1~mm thick windows, acting as secondary Cherenkov radiators, ensured Cherenkov photon emission starting from momenta lower than 1~GeV/$c$ for all the particle species of interest, allowing charged particle timing performance studies.

The S13552 arrays were bonded on one side of custom PCBs, with the 128 channels of each array arranged into 32 groups, each consisting of 4 adjacent channels, resulting in a $1\times1.625$ mm$^{2}$ readout pitch. Each of these groups of channels was routed to a Samtec LSHM-120 multi-channel connector bonded on the opposite side of the PCBs~\cite{MAZZIOTTA2022167040}.
The S13361-3075AE-08 arrays were instead plugged with their Samtec ST4-40 connectors to different custom PCBs with Samtec SS4-40 connectors and all channels were routed to two Samtec LSHM-120 multi-channel connectors bonded on the opposite side.
Custom FEBs equipped with four 32-channel Petiroc~2A front-end (FE) ASICs (37 ps LSB TDC, 10 bit ADC)~\cite{Fleury:2014hfa} were used to measure the signal charge and the arrival time. Each HLCD with 32 analog signals was read-out with a single Petiroc 2A ASIC. Each FEB hosts a CAEN A7585D SiPM voltage module~\cite{a7585} and a Kintex-7 FPGA mounted on a Mercury+ KX2 module~\cite{fpgakx2}. 
We have used four Petiroc 2A FEBs with one as controller providing to the others boards a common 40 MHz clock, trigger signal and event tag to synchronize them~\cite{MAZZIOTTA2022167040,Roberta_proceeding_IWASI}.

In the second version of the prototype (2024) the Cherenkov-photon detector was upgraded using HPK SiPM S13361-2050AE-08 arrays (8$\times$8 channels with 2.2 mm of pitch) to detect the Cherenkov photons produced in the aerogel radiator with centers arranged along a circumference of 5.9 cm of radius. A HPK S13361-3075AE-08 SiPM array (M2) with 1 mm SiO${_2}$ window was placed in the center along the beam line to perform charged particle timing measurement as in the first prototype~\cite{Mazziotta_IWORID}.
An additional cylindrical vessel placed upstream with respect to the one including the aerogel tile was assembled and instrumented with two HPK SiPM S13361-2050AE-08 arrays where we glued 1 mm thick SiO$_{2}$ window for particle timing studies. Those arrays were plugged with their Samtec ST4-20 connectors to a custom PCB with Samtec SS4-20 connectors and all channels were routed to two Samtec LSHM-120 connectors bonded on the opposite side. 
The electronics was upgraded using new custom FEBs equipped with one 64-channel Radioroc 2 FE ASICs developed by Weeroc~\cite{Saleem:2023pwt}, providing a much better response to single photo-electron signals than Petiroc 2A ASICs, coupled with the picoTDC ASIC developed by CERN~\cite{Altruda:2023qoh} to measure the arrival-time (ToA) and the time-over-threshold (ToT) of the analog SiPM signals.
Each HLCD with 32 analog signals was read-out with half of a Radioroc 2 and a picoTDC. The Radioroc2/picoTDC board was plugged to a MOSAIC read-out board~\cite{DeRobertis:2018vls}. The SiPM analog signals were amplified and discriminated with the Radioroc 2 FE (the threshold is set to a single photo-electron) and the digital signals are sent to the picoTDC to measure the ToA and the ToT, providing information on the SiPM number of photo-electrons. The picoTDC was configured with ToA LSB of 3.05 ps, ToT LSB of 195 ps and the acquisition window set to 200 ns.

\begin{figure}
\centering
\includegraphics[height=0.16\textheight]{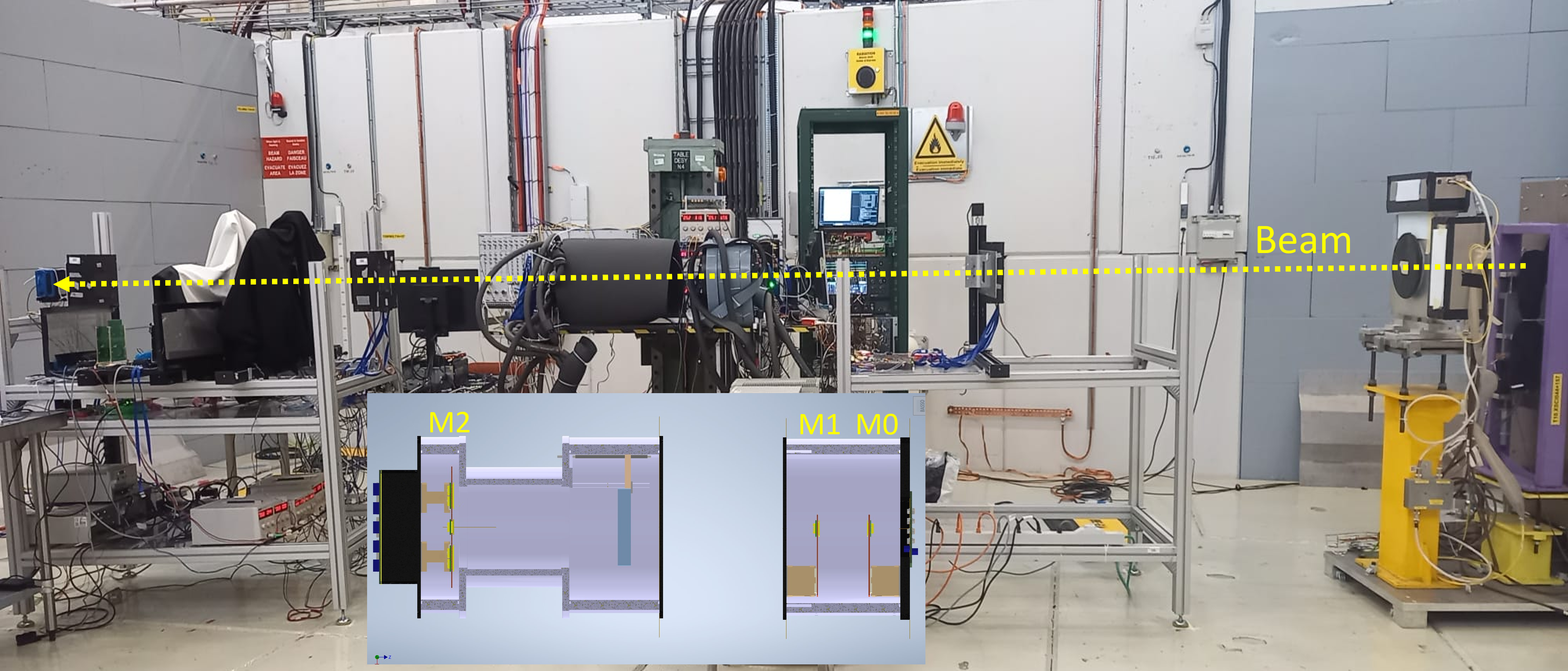}
\includegraphics[height=0.16\textheight]{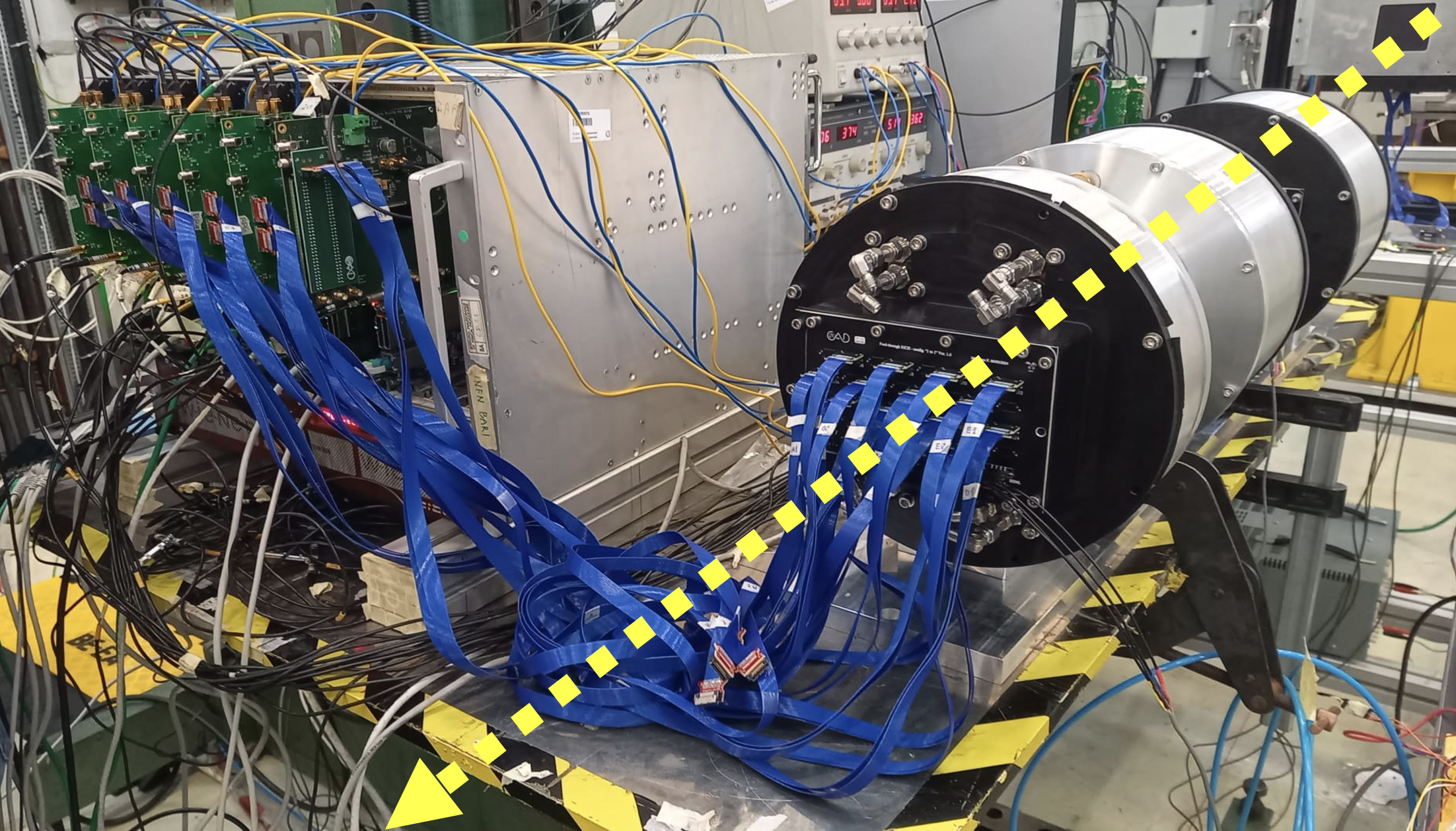}
\caption{Left: Beam test set-up at CERN PS T10 line on Sep/Oct, 2024. The beam enters from the right side. The two insets show and artist-view of the  RICH prototype inside the cylinders. The black box in the set-up includes a X-Y fiber tracker module~\cite{MAZZIOTTA2022167040}. In the 2023 set-up, only one cylinder was used including the aerogel tile and two arrays along the beam line \cite{Mazziotta_IWORID}. Right: A Crate housing the Radioroc2/picoTDC FEBs, each plugged on a MOSAIC board. The Samtec HLCD-20 cables are also shown to route the signals from the feed-through boards to the FEBs (see text for more details).}
\label{fig:setup}
\end{figure}

The set-up used in the 2024 at the CERN-PS T10 beam line is shown in Fig.~\ref{fig:setup}.
The vessels housing the tested RICH prototype and timing arrays are in the middle of the set-up. The upstream and downstream X-Y fiber tracker modules for beam particle triggering and tracking are also shown.
The right panel shows the connections to the electronics board as discussed above.

\section{Beam test results}
\label{sec:beamtest}

The beam test data analysis procedure is similar to the one discussed in Ref.s~\cite{Mazziotta_IWORID,rich1}. 
We assume all Cherenkov hits in the ring arrays as candidate Cherenkov photons produced in the aerogel tile by the impinging charged track. For the angle reconstruction we assume emission in the middle plane of the aerogel tile (corresponding to a 1 cm depth), with the position in the transverse plane being evaluated from the reconstructed track with the fiber tracker modules.
The corresponding hypothetical Cherenkov emission angle is then calculated according to the hit position, with a geometric backpropagation procedure to the emission point, also accounting for the different refractive indeices of the traversed media.

\begin{figure}
\centering
\includegraphics[width=0.75\linewidth,height=0.3\textheight]{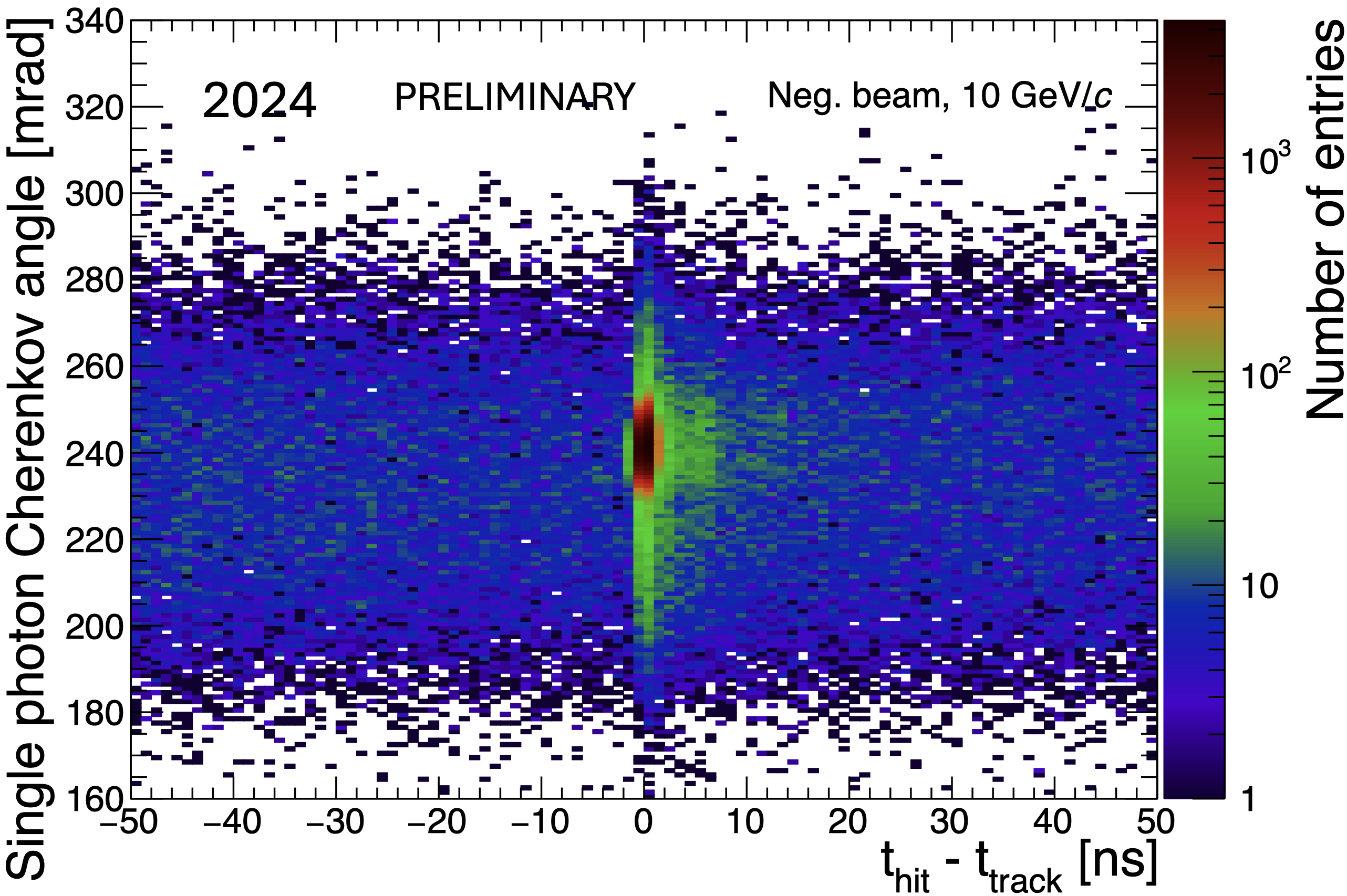}
\vspace{0.2cm}\\
\includegraphics[width=0.48\linewidth]{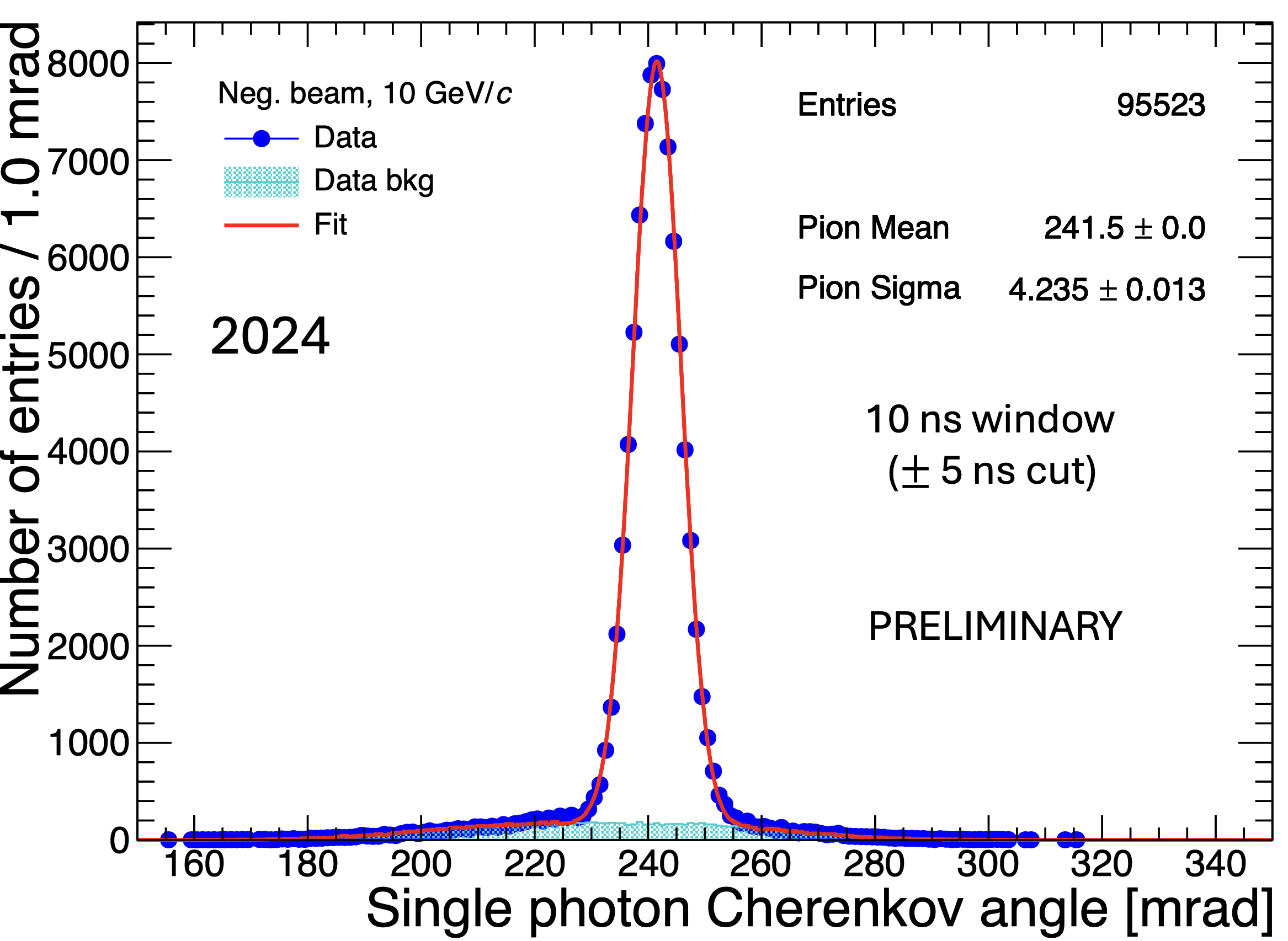}
\hspace{0.2cm}
\includegraphics[width=0.48\linewidth]{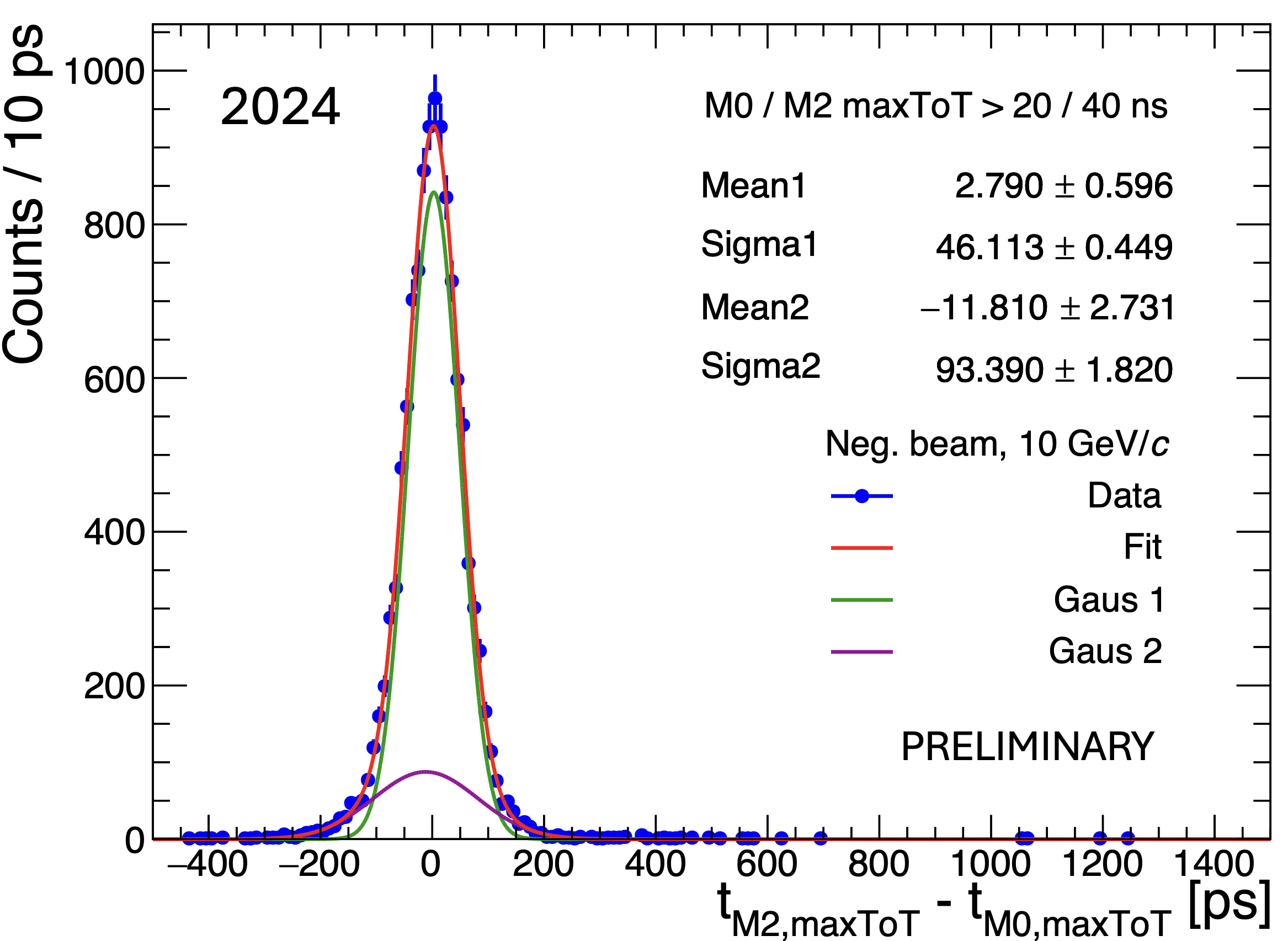}
\caption{Test beam results achieved in 2024 for the negative charged beam at 10 GeV/$\it{c}$ (mainly pions). Top: Cherenkov angle reconstruction as a function of the photon arrival times respect to the charged particle; Left bottom: Cherenkov angle distribution in a 10 ns time window; right bottom: M2-M0 timing performance.}
\label{fig:results_2024}
\end{figure}

Fig.~\ref{fig:results_2024} shows a summary of the results achieved with the data collected in the 2024 beam test campaign using the negative charged beam at 10 GeV/$\it{c}$ (mainly pions).
The top panel shows the distribution of the reconstructed Cherenkov angle $\theta_{ch}$ as a function of the photon arrival times with respect to the charged particle time measured in the M2 central SiPM array $\Delta t$. The sparse uniformly distributed hits corresponds to the uncorrelated background due to the electronics noise and the SiPM dark counts.
The vertical and horizontal bands superimposed to the expected pion cluster at $\theta_{ch}=241.5$~mrad and $\Delta t=0$~ns correspond to the correlated background hits due to Cherenkov photons undergoing Rayleigh scattering before leaving the aerogel and possible SiPM afterpulses or cross-talk effects, respectively.

The left bottom panel of the Fig.~\ref{fig:results_2024} shows the single photon Cherenkov angle distribution obtained selecting candidate Cherenkov hits with $|\Delta t|<5$~ns.
A significant suppression of the contribution of uncorrelated background hits is achieved thanks to the time cut applied in the hit selection.
The expected peak due to Cherenkov photons produced by pions in the areogel radiator is clearly visible.
The angle distribution is fitted with the sum of a Gaussian distribution corresponding to the signal peak and a template background distribution evaluated selecting only hits with $|\Delta t|>50$~ns.
The resulting angular resolution (in sigma unit) is about 4.2 mrad slightly worse than the one measured in 2023~\cite{Mazziotta_IWORID} due to the slightly larger SiPM 2D pitch (the ALICE 3 bRICH baseline $2.2\times2.2$ mm$^{2}$ of the 2024 prototype against the available $1 \times 1.625$~mm$^{2}$ of 2023 prototype). 

Finally, the right bottom panel of the Fig.~\ref{fig:results_2024} shows the distribution of the time difference between the two pixels with maximum picoTDC ToT in M2 and M0 requiring ToTs larger than 40~ns and 20~ns, respectively, and correcting for any time offset and the time walk effect (signal with the same shape and different amplitude cross the threshold at different times) as well.
Large ToT cut values allow us to select events with lower time jitter. 
The measured time distribution has a full width half maximum (FWHM) value of about 112 ps. The distribution is then fitted
with the sum of two Gaussian distributions corresponding to the core of the signal peak and its tail. 
The sigma of the core is about 46 ps on the difference between the times in the two arrays, that corresponds to about 46~ps/$\sqrt{2}$ $\approx$ 32~ps time resolution on average at the single SiPM level assuming the contributions of the two arrays to be equal, while the sigma of the tail is about 94 ps.
The larger value with respect to the core may be due to events with wrongly reconstructed timing, inaccuracies in time walk corrections (the measured time is corrected with the ToT), or limitations associated with the Radioroc2/picoTDC boards, which measure only the ToT rather than the charge. Additionally, the effect may be influenced by SiPM afterpulsing or impedance mismatches introduced by the 1-meter-long cable connecting the SiPM to the front-end electronics.

\section{Conclusions}
\label{sec:conc}

The results achieved with the tested prototypes are consistent with ALICE 3 bRICH specifications. 
We measured a single photon angular resolution at the Cherenkov angle saturation of 4.2 mrad using a 2~cm thick aerogel radiator having a refractive index of $1.03$.
We also achieved an excellent suppression of uncorrelated background hits using a short hit-track time matching window.
The measured resolution on the relative time between the charged particle hits in two different arrays placed along the beam line was $\approx$ 46 ps in sigma, corresponding to about 30 ps at single SiPM channel,  accounting both the intrinsic SiPM contributions and the electronics.

\section*{Acknowledgments}
The authors would like to thank the INFN Bari staff for its contribution to the procurement and to the construction of the prototype. In particular, we thank D. Dell'Olio, M. Franco, N. Lacalamita, F. Maiorano, M. Mongelli, M. Papagni, C. Pastore and R. Triggiani for their technical support.
The authors would like to thank Weeroc for contributing to the development of the Radioroc~2/picoTDC board and providing support for the operation of Radioroc~2.
The authors also acknowledge the CERN team for providing the facilities and support throughout all the beam test duration.


\bibliographystyle{unsrt}
\bibliography{ProceedingPD2024.bib}


\end{document}